\documentclass[aip,jcp,
numerical,
amsmath,amssymb,
reprint
]{revtex4-1}

\usepackage{amsmath}
\usepackage{xspace}
\usepackage{relsize}
\usepackage{tikz}
\usepackage[utf8]{inputenc}
\usepackage{bm} 
\usepackage{graphicx}
\usepackage{tabularx}
\usepackage{lpic}
\usepackage[tight]{units}
\usepackage[version=3]{mhchem}

\newcommand{\avery}     {{James Emil Avery}\xspace}
\newcommand{\averymail} {{avery@nbi.ku.dk}}
\newcommand{\dreuw}     {{Andreas Dreuw}\xspace}
\newcommand{\dreuwmail} {{dreuw@uni-heidelberg.de}}
\newcommand{\maxsch}    {{Maximilian Scheurer}\xspace}
\newcommand{\ademp}     {{Adrian L. Dempwolff}\xspace}
\newcommand{\mfh}       {{Michael F.~Herbst}\xspace}
\newcommand{\mfhmail}   {{michael.herbst@iwr.uni-heidelberg.de}}
\newcommand{\IWRaddress}{Interdisciplinary Center for Scientific Computing,
Heidelberg University, Im Neuenheimer Feld 205, 69120 Heidelberg, Germany}
\newcommand{\NBIaddress}{Niels Bohr Institute, University of Copenhagen,
Blegdamsvej 17, 2100 København, Denmark}

\newcommand{\python}  {\texttt{python}\xspace}
\newcommand{\cpp}     {\texttt{C++}\xspace}
\newcommand{\ccc}     {\texttt{C}\xspace}
\newcommand{\fortran} {\texttt{FORTRAN}\xspace}

\newcommand{\adcman}  {\texttt{adcman}\xspace}
\newcommand{\pyscf}   {\texttt{pyscf}\xspace}
\newcommand{\libint}  {\texttt{libint}\xspace}
\newcommand{\libcint} {\texttt{libcint}\xspace}
\newcommand{\numpy}   {\texttt{numpy}\xspace}

\newcommand{\molsturm}{\texttt{molsturm}\xspace}
\newcommand{\gscf}    {\texttt{gscf}\xspace}
\newcommand{\gint}    {\texttt{gint}\xspace}
\newcommand{\sturmint}{\texttt{sturmint}\xspace}
\newcommand{\lazyten} {\texttt{lazyten}\xspace}
\newcommand{\krims}   {\texttt{krims}\xspace}
\newcommand{\contraction}{contraction\xspace}

\newcommand{\ie}{\mbox{i.e.}\xspace}
\newcommand{\eg}{\mbox{e.g.}\xspace}

\renewcommand*{\vec}[1]{\ensuremath{\underline{\boldsymbol{#1}}}}
\newcommand*{\uvec}[1]{\ensuremath{\hat{\underline{\boldsymbol{#1}}}}}

\newcommand{\SCF}{SCF\xspace}
\newcommand{\HF}{HF\xspace}
\newcommand{\DFT}{DFT\xspace}
\newcommand{\ADC}{ADC\xspace}

\newcommand{\mat}[1]{\mathbf{#1}}

\begin{document}

\title{
	Towards quantum-chemical method development
	for arbitrary basis functions
}

\author{\mfh}
\email\mfhmail
\affiliation{\IWRaddress}

\author{\dreuw}
\email\dreuwmail
\affiliation{\IWRaddress}

\author{\avery}
\email\averymail
\affiliation{\NBIaddress}

\begin{abstract}
	We present the design of a flexible quantum-chemical
	method development framework,
	which supports 
	employing any type of basis function.
	This design has been implemented in the
	light-weight program package \molsturm,
	yielding a basis-function-independent self-consistent field scheme.
	Versatile interfaces,
	making use of open standards like \python,
	mediate the integration of \molsturm with existing third-party packages.
	In this way both rapid extension of the present set of
	methods for electronic structure calculations
	as well as adding new basis function types
	can be readily achieved.
	This makes \molsturm well-suitable for testing
	novel approaches for discretising the electronic wave function
	and allows comparing them
	to existing methods using the same software stack.
	This is illustrated by two examples,
	an implementation of coupled-cluster doubles
	as well as a gradient-free geometry optimisation,
	where in both cases,
	an arbitrary basis functions could be used.
	\molsturm is open-source and can be obtained from \url{http://molsturm.org}.
\end{abstract}
\keywords{electronic structure theory, method development, basis-function independence, molsturm}

\maketitle

\section{Introduction}
\label{sec:intro}
The central goal of electronic-structure theory
is to find approximate solutions to the electronic
wave equation numerically.
This requires a discretisation of
the electronic wave function.
Typically, it is approximated as
a linear combination of Slater determinants:
anti-symmetrised products of single-particle functions.
The latter are in turn constructed
by expanding them in a basis set of a priori determined
single-electron functions.
Usually, such basis sets are not complete
and introduce basis set errors.
Proper choice of the basis function type and size
of basis set is thus decisive for an accurate
description of the system under investigation.
It is also clear from the onset that different basis function types
can be more or less suited for a specific problem,
suggesting to conduct investigations across existing basis function types.

Gaussian-based methods are overwhelmingly predominant in
computational electronic structure theory,
which stems from pragmatic reasons dating back to the founding
years~\cite{Boys1950,Hehre1969}.
It was well-known that bound
state electronic wave functions decay exponentially both at short and
large distances from the nuclei~\cite{Morgan1977},
but multi-centre
electron-repulsion integrals (ERI) of products of exponential-type orbitals
were impractically difficult to calculate.
For Gaussian-type orbitals (GTO), on the other hand,
ERI could be calculated
efficiently due to the Gaussian Product Theorem.
However,
the computational challenges facing quantum chemists have changed
since the 1970's, and it may now be worth trading extra computation
per integral for having fewer, more accurate basis functions.

In many practical applications, the shortcomings of GTOs are
not important, or one is able to compensate
by employing specialised contracted basis sets~\cite{Jensen2013,Hill2013}.
However,
even contracted GTOs (cGTO) fail to describe both the nuclear cusp and
the exponential decay of the electron density~\cite{Kato1957}. In addition,
no matter the number of GTO basis functions used,
the derivatives are always wrong at the nucleus, which causes singularities
and computational failure for example in
quantum Monte Carlo calculations~\cite{Foulkes2001,Ma2005,Loos2017}.
Furthermore the description of some properties
such as the nuclear-magnetic resonance (NMR) shielding tensors
or a description of Rydberg-like or auto-ionising states%
~\cite{Feshbach1958,Feshbach1962,Riss1993,Santra2002}
directly involves the nuclear cusp or the asymptotic tail,
making physically accurate basis functions desirable~\cite{Guell2008,Hoggan2009}.

The name of our implementation, ``\molsturm'',
is a portmanteau of \emph{molecular Sturmians}: the project
was born as a means to solve the problem of using state-of-the-art quantum chemistry methods
together with generalised- and molecular Sturmian basis functions.
The many promising results for generalised Sturmians were stranded due to the fact that
only electronic structure problems that were small enough to be solved by direct 
configuration interaction methods could be treated, preventing wider use.
The existing mature quantum chemistry software has been developed over 
hundreds of man-years, and the methods are not easily reimplemented from scratch.

In theory, it should be a simple matter to include new basis function
types in existing quantum chemistry software by swapping the integral
calculator. In practice, it turned out to be exceedingly difficult
due to the very large and complicated code bases of all the
investigated quantum chemistry software. Assumptions about the basis
function type scattered around the source code only make this even
task more difficult.

Our solution, which is presented in this paper, is to implement a
light-weight layer that makes it easy to experiment with
many different basis function types and quantum-chemical methods.
It is designed for researchers to both build simple stand-alone
programs for prototyping and teaching purposes, and to make plug-in
modules to be hosted in standard quantum chemistry software.
To the best of our knowledge such a framework with the ability
to explore quantum-chemical methods across multiple basis function types
has been missing up to today.

\subsection{Alternative Basis Function Types}
Many research groups have worked on alternative basis function types.
This section will provide a brief overview with particular focus
on exponential-type orbitals~(ETO).
For further details regarding the basis function mentioned,
the reader is referred to the cited works.

Efforts on making various types of ETOs
computationally feasible were pioneered
by Harris, Michels, Steinborn, Weniger, Weatherford, Jones, and others%
~\cite{HarrisMichels1967,Steinborn1983,Weniger1983,Weatherford1982}.
A particular form of complete ETO basis are Coulomb Sturmians~\cite{Shull1959,Rotenberg1970,Aquilanti2003,Coletti2013,Calderini2012}~(CS).
Their functional form is identical to the familiar hydrogen-like orbitals,
just with all occurrences of the factors $Z/r$ replaced by
the Sturmian exponent $k$ --- a parameter,
which is the same for all functions of the basis:
\begin{equation}
	\varphi^\text{CS}_{nlm}(\vec{r}) =
	k^{3/2} N_{nl} (2k r)^l e^{-k r}
	L^{2l+1}_{n-l-1}(2k r) Y_l^m(\uvec{r}).
	\label{eqn:CSdef}
\end{equation}
In this, $Y_l^m$ is a spherical harmonic,
$L^{2l+1}_{n-l-1}$ an associated Laguerre polynomial and
\begin{equation}
N_{nl} = \frac{2}{(2l+1)!} \sqrt{ \frac{(l+n)!}{n (n-l-1)!}}
\end{equation}
a normalisation constant.
CS functions proved to be especially easy to work with,
since their momentum-space
representation by hyperspherical harmonics allows efficient
calculation of multi-centre integrals, opening the way for 
efficient molecular calculations~\cite{Avery2012,Avery2013,Avery2015,Avery2017}.
The Coulomb Sturmian construction generalises well
and generalised Sturmian basis sets preserving many useful Sturmian properties
can be constructed.
These allow, for example, to build $N$-particle basis functions
that include important
geometric properties of the physical system under consideration
at the level of the basis%
~\cite{Aquilanti1996,Aquilanti1997,Aquilanti1998,Avery2003,Avery2004,Avery2006,Avery2009,Calderini2013,Abdouraman2016}.
Similarly, $d$-dimensional hyperspherical harmonic basis sets can model collective
motions of particles, for example for treating strongly interacting few-body systems 
or reactive scattering~\cite{Avery1989-hyperbook,Aquilanti1992,Aquilanti2004,Avery2018-hyperbook,Das2016}.
A particular type of one-particle Sturmians
combines a bound-state region and plane-wave asymptotics to model photoionisation
in scattering~\cite{Randazzo2010,Mitnik2011,Randazzo2015,Granados2016}.
Other directions of research
towards alternative discretisation methods
include quantum chemistry on numerical real-space grids~\cite{Soler2002,Frediani2015},
finite element methods%
~\cite{Tsuchida1995,Lehtovaara2009,Alizadegan2010,Avery2011Phd,Davydov2015,Boffi2016},
and wavelets~\cite{Bischoff2011,Bischoff2012,Bischoff2013,Bischoff2014,Bischoff2014a,Bischoff2017}.
%

The \molsturm package is to support such research directions
by providing a common platform for development, testing and analysis of
quantum-chemical methods irrespective of the basis function type
employed for the discretisation. The goal is for the implementation
work of introducing a new basis function type to be
reduced to adding an extra integral back end in \molsturm, which will
then both provide simple stand-alone calculations and a common interface
to hook into existing quantum chemistry packages.

\subsection{Towards Basis-type Agnostic Quantum Chemistry}
In order to reach a basis-type agnostic design,
there are three fundamental components to consider:
(i) an integral interface accommodating a wide range of very different
basis set types and discretisation, but providing a uniform way of
accessing them, (ii) simple discretisation-agnostic implementations
of the self-consistent field (SCF) algorithms,
and (iii) a flexible interface to employ the resulting SCF orbital
basis further in existing, third-party code.
Once the SCF orbitals have been obtained,
the remainder of a calculation, \eg a Post-Hartree-Fock~(Post-HF) method,
can usually be formulated entirely in the SCF orbital basis,
without reference to the underlying basis functions.
Thus, a basis-function independent SCF scheme automatically
leads to basis-function independent Post-HF methods as well.

This structure has another advantageous side effect in the context
of developing new basis function types,
as it allows to perform comparisons between old and new methods
using exactly the same software stack.
In other words one can thus be sure
that, apart from the discretisation,
all aspects of the calculation
\eg SCF algorithms or guess methods, 
are optimised at the same level
leading to a fair apples-to-apples comparison
between old and new methods.

\subsection{Paper Outline}
The remainder of the paper is structured as follows:
Section \ref{sec:related} reviews
existing projects with similar goals to \molsturm.
Section \ref{sec:theory} provides a theoretical background for
the program design choices, which are described in Section \ref{sec:design}.
Section \ref{sec:examples} provides example problems calculated using
\molsturm's \python interface, illustrating
how to implement new methods on top of \molsturm in a few lines of \python.
Section \ref{sec:state} outlines the current state of
\molsturm and what we hope to achieve in the future.


\section{Related quantum-chemical software packages}
\newcommand{\psifour}{\texttt{Psi4}\xspace}
\newcommand{\pyquante}{\texttt{pyQuante}\xspace}
\newcommand{\horton}{\texttt{HORTON}\xspace}
\newcommand{\gpaw}{\texttt{GPAW}\xspace}
\newcommand{\ASE}{\texttt{ASE}\xspace}
\newcommand{\CPtK}{\texttt{CP2K}\xspace}
\newcommand{\psifnp}{\texttt{Psi4NumPy}\xspace}

\label{sec:related}
%

This section reviews existing quantum chemistry software that share
some of the goals of \molsturm.

The quantum Monte Carlo packages CASINO~\cite{Needs2010} and QMCPACK~\cite{Kim2012}
are among the few systems that support many different basis function types.
Both programs support discretizations
in terms of GTOs, STOs, plane-waves, and numerical orbitals like splines.
Similarly, the packages \CPtK~\cite{Hutter2014}, \ASE~\cite{Larsen2017}
and \gpaw~\cite{Mortensen2005,Enkovaara2010}
can be employed to perform and post-process computations
using more than one type of basis function.
\gpaw and \CPtK further support calculations
with hybrid basis sets that mix
Gaussian-type orbitals with plane waves.
However, to the best of our knowledge,
the design of these packages is
very specific to the particular combinations of basis function type
and method.

Combining a \fortran or \ccc/\cpp
implementation of the time-critical core with \python as a
high-level interface language works exceedingly well, 
and this solution has become increasingly popular.
\citet{Sun2017}~describe the reasons as follows in their paper about \pyscf:
\begin{itemize}
	\item There is no need to learn a particular domain-specific
		input format.
	\item All language elements from \python are immediately
		available to \eg automatise repetitive calculations
		with loops or similar.
	\item The code is easily extensible beyond what is available
		inside \pyscf, for example to facilitate plotting
		or other kinds of analysis.
	\item Computations can be done interactively,
		which is helpful for testing or debugging.
\end{itemize}
We add here
that \python as a high-productivity language
often achieves even complicated tasks with few lines of code
while remaining easy to read and understand, demonstrated for
example by the coupled cluster implementation shown in Section~\ref{sec:examples}.
In the context of quantum chemistry
this has the pleasant side effect that a \python script
used for performing calculations and subsequent analysis
is typically brief,
but still documents the exact procedure followed.
All this comes at essentially
no downside
if \python is combined with
carefully optimised low-level \ccc or \fortran
code in the numerical hot spots.
\citet{Sun2017} for example claim that \pyscf is as
fast as any other existing quantum chemistry packages
written solely in \ccc or \fortran.

Even meta-projects like \ASE~\cite{Larsen2017} or \texttt{cclib}~\cite{Cclib}
which aim at extending existing packages by a common \python front end,
have emerged.
Other packages like \horton~\cite{Verstraelen2017}, \pyscf~\cite{Sun2017},
\pyquante~\cite{PyQuante} and \gpaw~\cite{Mortensen2005,Enkovaara2010} are written
almost exclusively in \python and only employ low-level \ccc or \cpp
code for the computationally demanding routines to various extents.

Starting from the opposite direction \psifour~\cite{Parrish2017} has
gradually introduced a more and more powerful \python interface on top of
their existing \cpp core over the years.
%
Recently their efforts have lead to the \psifnp project~\cite{Smith2018},
which combines the \python interface of \psifour
with the tensor operation syntax of \numpy arrays~\cite{Walt2011}.
The aims of \psifnp are very much in line with \molsturm,
namely to provide a framework,
which yields flexible and easy-to-read codes.
It is thus highly suitable for reference implementations, rapid prototyping
or teaching\cite{Smith2018}.
Unlike \molsturm, however, \psifnp
does not exhibit a basis-type agnostic design
and only supports discretisations
based on cGTO basis sets.

Another common feature to \pyscf and \psifour is their modular design.
They use well-established open standards
like HDF5~\cite{HDF5Manual} or \numpy arrays~\cite{Walt2011}
for data exchange,
such that linking their codes to external projects becomes easy.
\psifour for example managed to integrate more than 15 external packages
into their framework.
This includes three completely different back ends for the computation of the
required integrals.
In the case of \pyscf it only took us about a day to link
our \molsturm to the full configuration interaction~(FCI) algorithms of \pyscf
via an interface based on \numpy.
Nevertheless the numerical requirements of Gaussian-type orbitals
are currently hard-coded inside the optimised
\ccc or \cpp parts of both these projects,
such that extending them by other types of basis functions could still be difficult.


\section{Theory}
\label{sec:theory}

This section briefly discusses the theoretical background
and properties of self-consistent field problems in the context
of the basis-type independent design aspired for \molsturm.
A more detailed analysis is provided in reference \onlinecite{Herbst2018Phd}.

\subsection{Self-Consistent Field Schemes}
\label{sec:th-scf}



Both Hartree-Fock~(\HF)
as well as Kohn-Sham density-functional theory~(\DFT)
can be viewed as a minimisation procedure of an energy functional
with respect to the occupied \HF or \DFT orbitals%
~\cite{Cances2000,Cances2000a,Cances2000b}.
After employing a particular basis set for discretisation this
minimisation problem becomes parametrised in the orbital coefficients $\mat{C}$
and the associated Euler-Lagrange equations may be written as:
\begin{equation}
	\begin{aligned}
	\mat{F}\!\left(\mat{C}\right) \mat{C} &= \mat{S} \mat{C} \mat{E}, \\
	\mat{C}^\dagger \mat{S} \mat{C} &= \mat{I},
	\end{aligned}
	\label{eqn:SCFeulerLagrange}
\end{equation}
where $\mat{C}$ is the matrix of occupied orbital coefficients,
$\mat{S}$ is the overlap matrix, $\mat{I}$ is the identity matrix
and $\mat{E}$ is the diagonal matrix of orbital energies.
The Kohn-Sham or Fock matrix $\mat{F}\!\left( \mat{C} \right)$
itself depends on the solution coefficients $\mat{C}$,
making \eqref{eqn:SCFeulerLagrange} a non-linear eigenproblem.
In the following our focus will be on the
\HF problem, since \molsturm currently does not implement
any \DFT exchange-correlation functional.
Due to the structural similarity of both \HF and \DFT,
our approach nevertheless applies to \DFT as well.

Because \eqref{eqn:SCFeulerLagrange} is non-linear,
the \HF problem must be solved iteratively.
Starting from an initial guess $\mat{C}^{(0)}$,
the SCF procedure aims to construct a sequence of trial matrices
$\mat{C}^{(1)}, \mat{C}^{(2)}, \ldots, \mat{C}^{(n)}$
converging towards the minimiser of the \HF energy functional.
Broadly speaking this can be achieved in two ways,
either by directly minimising the \HF energy functional%
~\cite{McWeeny1956,Igawa1975,Seeger1976,Voorhis2002}
or alternatively by satisfying \eqref{eqn:SCFeulerLagrange}
~\cite{Roothaan1951,Saunders1973,Pulay1980,Pulay1982},
thus obtaining a stationary point on the \SCF manifold.
On top of that one may alternatively formulate the \HF problem,
such that instead of the coefficient matrix,
the density matrix
\begin{equation}
\mat{D}^{(n)} = \mat{C}^{(n)} \left(\mat{C}^{(n)}\right)^\dagger
\end{equation}
is iterated.
To distinguish \SCF algorithms according to these parametrisations,
we will refer to the latter kind of \SCF algorithms as
\textit{density-matrix-based SCF} schemes,
whereas we will use the term
\textit{coefficient-based SCF}
for the former set of algorithms~\cite{Herbst2018Phd}.

\begin{figure*}
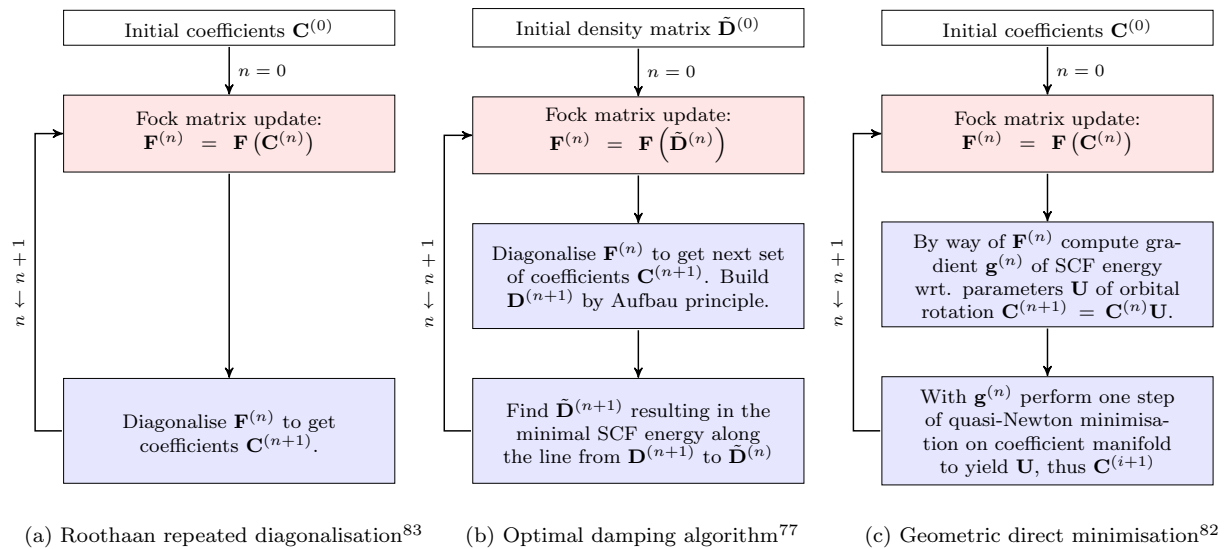

	\centering
	\begin{lpic}{scf_schemes(0.93)}
		\lbl{31.3,3;\smaller(a) Roothaan repeated diagonalisation~\cite{Roothaan1951}}
		\lbl{89.7,3;\smaller(b) Optimal damping algorithm~\cite{Cances2000a}}
		\lbl{148.3,3;\smaller(c) Geometric direct minimisation~\cite{Voorhis2002}}
	\end{lpic}%
	\\[0.4em]  
	\caption{Schematic overview of a few exemplary
		self-consistent field algorithms.
		In each case the Fock-update step is highlighted in pale red
		and the steps updating the coefficients or the density matrix
		are shaded in pale blue.
		For further details regarding the algorithms see the indicated references.
	}
	\label{fig:SCFs}
\end{figure*}
To illustrate, 
figure \ref{fig:SCFs} shows simplified schemes for three \SCF algorithms:
Roothaan's repeated diagonalisation~\cite{Roothaan1951},
the optimal damping algorithm~(ODA)~\cite{Cances2000a},
and the geometric direct minimisation~(GDM) scheme~\cite{Voorhis2002}.
While Roothaan's algorithm and the GDM are coefficient-based,
the ODA is density-matrix-based.
Roothaan's algorithm is the simplest representative
for solving \eqref{eqn:SCFeulerLagrange} by repetitively
treating the standard eigenproblem that arises from fixing
$\mat{F}\!\left( \mat{C}^{(n)} \right)$.
In contrast to this, the
GDM directly minimises the energy functional geometrically.
The ODA is a middle ground: It combines a line-search minimisation
of the energy with respect to the density matrix
with repeated diagonalisation of arising Fock matrices.

The aforementioned algorithms can --- on an abstract level ---
be written as a two-step process,
where a \textit{Fock-update} step and
a \textit{coefficient-update} or \textit{density-matrix-update}
step are iterated.
In the former step, a new Fock matrix
$\mat{F}^{(n)}$ is constructed from
the current set of \SCF coefficients $\mat{C}^{(n)}$
or the current density matrix $\mat{D}^{(n)}$ (red boxes in figure \ref{fig:SCFs}).
In the second step a new set of coefficients $\mat{C}^{(n+1)}$
or a new density matrix $\mat{D}^{(n+1)}$ is found
by means of the Fock matrix $\mat{F}^{(n)}$~(blue boxes in figure \ref{fig:SCFs}).
Typically other results obtained in previous iterations
are taken into account in this step as well to accelerate convergence.
Consider for example Pulay's commutator
direct inversion of the iterative subspace~(DIIS)~\cite{Pulay1982} scheme
forming a linear combination of previous Fock matrices.

Note, that
apart from the initial discretisation of the \HF or \DFT problem,
no reference to the basis function type was required in our discussion
about \SCF procedures.
In other words,
provided that (i) an \SCF algorithm can be brought into two-step form
and that (ii) within these steps the details of the basis function
can be hidden,
the algorithm can be implemented without making explicit
reference to the underlying basis.
We are unaware of an \SCF algorithm which cannot be written in two-step form
and will henceforth concentrate primarily on the second point in our discussion
towards a basis-type-independent \SCF scheme.

\subsection{Matrix Structure and Contraction-based Methods}
\label{sec:th-contraction}
Because different basis types can
have very different selection rules and other discretisation properties,
the structure of the Fock matrix --- as well as the
numerical approaches required to efficiently solve the \SCF problem --- may vary.
This in turn affects the requirements
we need for the interface to the update steps,
which is a challenge for hiding the basis function details from the \SCF.
This subsection briefly discusses contraction-based methods as a solution to this issue.

\begin{figure*}
	\centering
	\includegraphics[width=\textwidth]{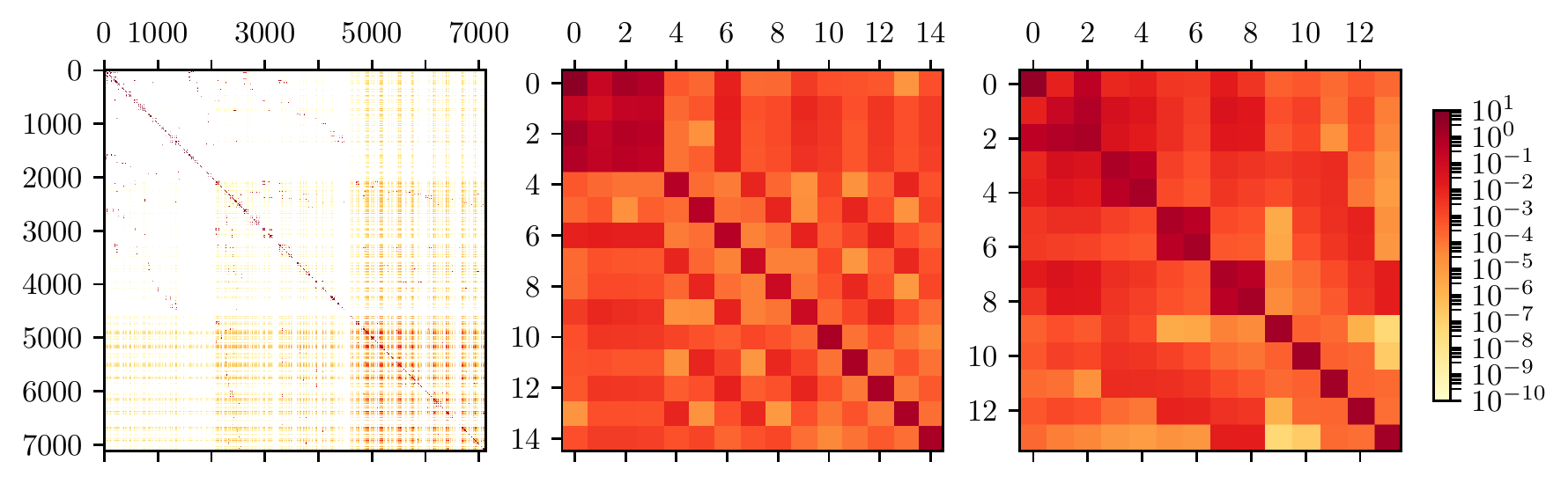}  \\[-0.5em]
	\begin{tabularx}{\textwidth}{
			@{\hspace{0.8cm}}>{\centering\arraybackslash}X
			>{\centering\arraybackslash}X
			>{\centering\arraybackslash}X@{\hspace{1.5cm}}
		}
		(a) $Q_2$ finite elements, &
		(b) contracted Gaussians, &
		(c) Coulomb Sturmians, \\
		adaptively refined 3D grid &
		pc-2 basis set~\cite{Jensen2007} & (3,2,2) basis with $k = 2.0$
	\end{tabularx}
	\caption{
		Structure of the Fock matrix
		at the beginning of a
		Hartree-Fock SCF calculation of beryllium,
		discretised using finite elements, contracted Gaussians, and Coulomb Sturmians.
		The 
		elements are coloured by the $\log_{10}$-scale shown on the right.
		The $(3,2,2)$ CS basis set of calculation (c)
		contains all CS functions \eqref{eqn:CSdef}
		whose quantum numbers $n, l, m$ satisfy
		$n \leq 3$, $l \leq 2$ and $m \leq 2$
		and with the exponent parameter chosen as $k = 2.0$.
		More details about construction schemes for CS basis sets
		can be found in reference \onlinecite{Herbst2018Phd}.
	}
	\label{fig:FockStructure}
\end{figure*}
Figure \ref{fig:FockStructure} shows from left to right
the Fock matrices arising if
(a) finite elements, piece-wise polynomials on a real-space grid%
~\cite{Grossmann1992,Brenner2008},
(b) contracted Gaussians or (c) Coulomb Sturmians have been employed
as the basis.
While the
Coulomb Sturmian and Gaussian matrices are both
small, dense and diagonal dominant,
the finite-element-discretised matrix
is sparse, but also much larger.
In fact, for a description of the beryllium atom density
at a relevant accuracy even larger basis sets with
$10^5$ to $10^6$ finite elements
are needed~\cite{Davydov2015}.

As a result, for both cGTO as well as CS-based discretisations,
direct diagonalisation algorithms,
\ie, where the full Fock matrix is diagonalised completely,
are applicable.
For finite-element-based approaches, on the other hand,
iterative subspace-based algorithms like Arnoldi's method~\cite{Arnoldi1951} or
Davidson's method~\cite{Davidson1975} are more common, due to the size of the matrix.
Recently these have been combined with
so-called matrix-free methods~\cite{Kronbichler2012},
which avoid building the finite-element problem matrix in memory at all.
Instead only an \emph{expression} for the
computation of the matrix-vector product is passed to an iterative solver.
Since such expressions may in general involve arbitrary tensor contractions,
like a contraction over the ERI tensor to compute the
Coulomb or the exchange part of the Fock matrix,
we will refer to these approaches by the term \textit{contraction-based methods}~\cite{Herbst2018Phd}.

The main driving force for such approaches is usually to reduce the amount
of storage required and instead employ well-optimised, high-throughput
matrix-vector contraction expressions.
For cases where this avoids slow storage such as hard drives,
runtime may be reduced significantly even though
matrix data will effectively be computed over and over.
Examples for contraction-based methods
in electronic structure theory are efficient implementations of Post-HF methods,
like the algebraic diagrammatic construction~(\ADC)
scheme~\cite{Wormit2009,Wormit2014,Dreuw2014}
as well as modern coupled-cluster algorithms~\cite{Helgaker2013}.
One should mention that in this context the
contraction expressions are usually called \textit{working equations}.

Note, however, that
contraction-based methods may even be favourable for cases
where the size of the system matrix
allows to get around using the hard drive
and place it in main memory instead~\cite{Herbst2018Phd}.
This can be understood by considering modern hardware trends.
State-of-the-art central processing units can perform
on the order of 1000 floating point
operations during the time needed to load data from main memory~\cite{CpuData},
a number which is likely going to increase in the future~\cite{Gocon2014}.
This implies that more and more algorithms may become bound by memory latency and bandwidth
rather than computation.
Especially for cases where matrix elements are fast to compute
from smaller, stored intermediates or even from tractable analytic expressions,
contraction-based methods are highly suitable.
Additionally a contraction-based approach often allows
to reorder terms in the matrix-vector product
or make use of discretisation-specific properties
providing additional sources of reducing runtime cost.
Such advantageous side-effects apply not only
to finite-element-based \HF,
but to Coulomb-Sturmian-based Hartree-Fock%
~\cite{Kronbichler2012,Avery2015,Herbst2018Phd} as well,
making contraction-based methods worth considering
in a context where multiple basis functions ought to be employed.

Additionally, a contraction-based approach can be readily
combined with the two-step SCF described in the previous subsection.
Focusing on a coefficient-based SCF for a second,
a contraction-based ansatz would implement the Fock matrix
$\mat{F}\!\left( \mat{C} \right)$ as a matrix expression
with the current coefficients $\mat{C}$ as some mutable state~\cite{Herbst2018Phd}.
The Fock-update step then amounts to transparently replacing
the current $\mat{C}$ in the Fock matrix expression,
which is a trivial process.
Furthermore, both ways to think about the \HF problem,
namely to think of it as a non-linear eigenproblem as well as an optimisation problem,
can be tackled by iteratively solving
appropriate linear problems or eigenproblems.
This can be achieved using a wide range of
iterative, subspace-based algorithms
such as GMRES, conjugate-gradient, Arnoldi's method or Davidson's method%
~\cite{Arnoldi1951,Saad2003,Arbenz2010,Saad2011,Davidson1975},
For a density-matrix-based SCF scheme, a
contraction-based formulation is possible as well.
Since the density matrix and the Fock matrix
have similar memory requirements and the density matrix
inevitably needs to be stored in a density-matrix-based SCF,
the main prospect of contraction-based methods,
to avoid the memory bottleneck of storing the Fock matrix,
is directly subverted.
Our discussion will not consider density-matrix-based SCF schemes
further for this reason.

In principle the Fock update step may be implemented by a conventional
re-computation of the full Fock matrix like in a cGTO setting.
Similarly, the matrix-vector-product expression may be realised
by multiplying the resulting stored matrix with an appropriate vector.
Thus the contraction-based SCF scheme
is a generalisation of the traditional method,
that provides additional flexibility
to deal with Fock matrices of various structures.
To conclude a single contraction-based interface between SCF algorithm and Fock matrix
is sufficient to provide a contraction-based SCF
irrespective of the basis function type and resulting matrix structures.

One should notice, however, that
iterative eigensolvers are not appropriate for all systems.
For some cGTO discretisations
with small and dense Fock matrices stored in memory,
direct solver methods perform better
than iterative ones and are thus preferable.
Optimal performance requires an abstraction layer
that --- depending on the basis function type and matrix structure ---
transparently switches
(i) between contraction expressions
and dense matrices for representing the Fock matrix,
and (ii) between iterative or direct solver algorithms.
As will be discussed in section \ref{sec:Lazyten}
the \lazyten lazy matrix library
is used for this purpose 
to achieve a basis-type-independent SCF code.
The details of the solver algorithm switching
and the basis-dependent routines for computation
are hidden in the abstraction layer of the linear algebra
and the contraction expression.


\section{Program Design}
\label{sec:design}
\subsection{Design Goals}
\label{sec:design-goals}

As mentioned above \molsturm aims to remove the
difficulty in implementing new types of basis functions and discretisations,
and to simplify
experimenting with new computational methods in quantum chemistry.
Assessment of new methods and comparison between old and new
should be possible within the same framework to ensure treatment
on an equal footing.
A high-level interface aiding automation
of repetitive calculations is desirable, too.
Once the trial phase is completed,
it should be easy to incorporate the new methods into existing quantum chemistry
software and thus make them widely available.
This motivates the overall design goals of \molsturm:

\paragraph{Enable rapid development.}
In the early stages of developing
a new quantum-chemical method, it is often not clear
how it will perform in practice or which approaches are required
to yield efficient and stable algorithms.
To simplify implementation, code should be high level
and close to the physical formulae,
and at the same time be flexible enough to
enable experimentation with different numerical methods.
Section~\ref{sec:python} discusses further details.

\paragraph{Plug-and-play implementation of new discretisations.}
It is a significant barrier to incorporating new basis function types
and discretisation schemes to quantum chemistry, that assumptions
about basis function types are scattered around in the --- often very
large --- programs.  We have designed \molsturm to isolate this to the
actual electronic integral back ends, and otherwise only where absolutely
necessary. The SCF stage and post-\HF methods should only
know about integrals on an abstract level.
Since symmetry, sparsity, selection
rules, and recursion rules are basis function dependent, the integral
library should be in charge of performing the operations where this
information is used. This is primarily in the tensor contractions, for
example the ERI-contractions with molecular coefficients.
Subsection~\ref{sec:GscfGint} discusses how this is done,
expanding on the strategies introduced in section~\ref{sec:theory}.

\paragraph{Easy interfacing with existing code.}

A challenge for new quantum-chemical methods is
that they are hard to compare to well-established ones:
One is either restricted to toy problems, or faced with the enormous
task of implementing advanced methods, refined over
hundreds of man-years in state-of-the-art quantum chemistry packages
--- clearly a rather daunting task.
For this reason, it is explicitly \emph{not} our goal to create yet another
general purpose quantum chemistry package and the large ecosystem of
functionality needed in such a project, but on the contrary to
supply small, flexible modules that can both be used on their own
for experimentation and teaching, and can be easily incorporated into
existing quantum chemistry software by simple interfaces.
For details see section \ref{sec:python}
as well as the examples in section \ref{sec:examples}.

\begin{figure*}
	\includegraphics[scale=1.1]{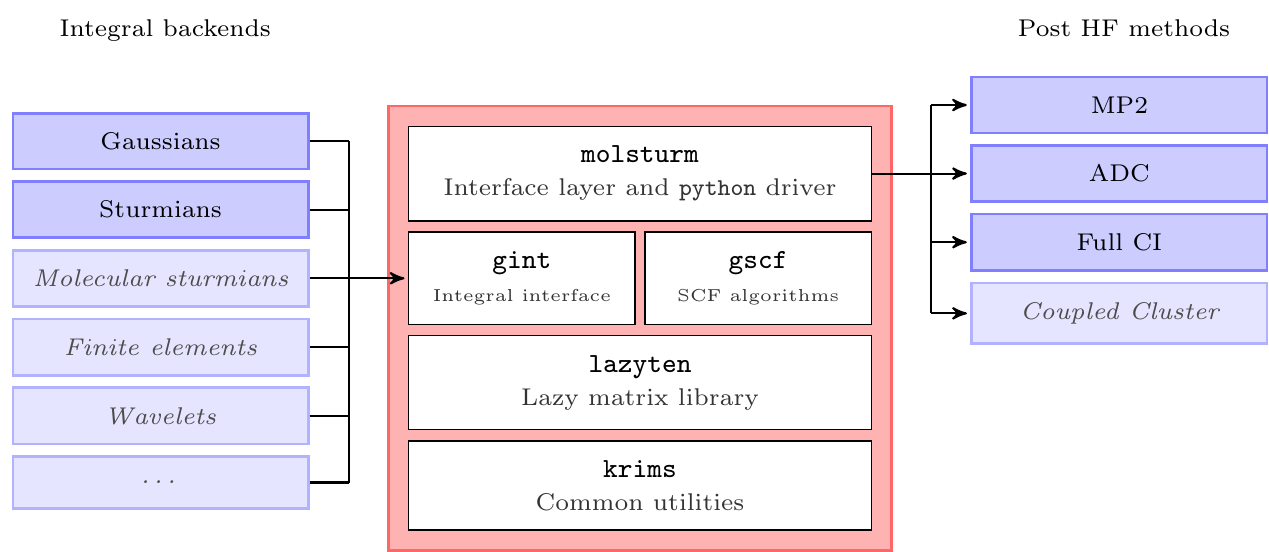}
	\caption{Structure of the \molsturm framework:
	Shown are the five modules of the package,
	along with third-party integral back ends and post-HF methods
	to indicate the mediator role of \molsturm.
	The greyed-out parts are not yet implemented,
	but could be supported by the design.}
	\label{fig:structure}
\end{figure*}

\paragraph{Modular structure with low code complexity.}
The aspired flexibility
requires that individual modules are as independent from another as possible.
We therefore choose a design in \molsturm,
where the five main modules are arranged in layers,
see figure \ref{fig:structure}.
Dependencies between the modules are only downwards, never sideways or upwards.
This aids both reuse of \molsturm's modules in external projects
as well as restructuring or replacement of code
if this was required in the future.
This is further aided by \molsturm's test suite,
which employs a range of testing strategies
including unit tests, functionality tests
as well as property-based tests.

Most \molsturm modules are written in \cpp, but
the top layer of the program is a \python module
defining the user interface of the framework.
Below this,
\gint, the \textit{general integral library},
provides a single link to multiplex between the supported integral calculation back ends
and \gscf implements the \contraction-based \SCF schemes,
following the general two-step update structure mentioned in section \ref{sec:th-scf}.
Both modules use the library \lazyten,
which defines a generalised linear algebra interface
allowing to transparently use dense, sparse and \contraction-based
Fock matrices in \gint and \gscf,
see section \ref{sec:Lazyten}.
Finally \krims is \molsturm's common utility library, named after the
German word ``\textit{Krims}krams'' for ``odds and ends''.
The individual components are discussed further in the next sections.

%
%
\subsection{Library for contraction-based algorithms}
\label{sec:Lazyten}

We saw in section \ref{sec:th-contraction} that
contraction-based methods provide a versatile
ansatz for self-consistent field algorithms,
leading to a basis-function independent formulation
of the problem.
We noted, that when the
Fock matrix is small and should be stored in memory,
an abstraction layer to switch between dense and iterative
solver schemes is needed for maximal efficiency.
On top of that an issue with contraction-based methods
is that the expressions for computing the matrix-vector products
can become complicated,
such that these are less intuitive to handle
compared to plain matrix or tensor operations.

Inside \molsturm, the library \lazyten solves these challenges by
representing matrices by a datastructure called a \emph{lazy matrix}\cite{Herbst2018Phd}.
These employ \emph{lazy evaluation}, a rigorous method from
programming language theory that allows postponing
computation until the moment it is needed.\cite{Hudak1989}

In contrast to a \textit{stored matrix},
which we define as a dense table, which has all
its elements residing in a continuous chunk of memory,
this restriction does no longer hold for a lazy matrix.
It may, for example, follow a particular sparse storage scheme
like a compressed-row format~\cite{Buluc2009},
but it may not even be associated to any kind of storage at all.
In the most general sense, it can be thought of as an
arbitrary contraction expression for computing the matrix elements,
which is dressed to look like an ordinary matrix from the outside.
One may still obtain individual matrix elements
and add, subtract or multiply lazy matrix objects, but
not all operations are as efficient
as for stored matrices.
Most importantly, accessing individual elements of lazy matrices
can be costly,
since the elements may be computed e.g.~from a particular
tensor contraction.
However, contraction operations of such objects must be fast.

Lazy matrix operations
are subject to \textit{lazy evaluation},
explaining the name of these data structures.
Lazy evaluation is a prominent concept of
functional programming languages, excellently introduced
in ref.~\onlinecite{Hughes1990}.
In our context, this means
that operations between lazy matrices
are not directly performed,
but delayed until a contraction of the resulting
expression with a vector or a stored matrix
unavoidably requires evaluation.
To illustrate this, consider the instructions
\begin{equation}
	\begin{aligned}
		\label{eqn:LazyMatrixInstructions}
		\mat{D} &= \mat{A} + \mat{B}, \\
		\mat{E} &= \mat{D} \mat{C}, \\
		\vec{y} &= \mat{E} \vec{x},
	\end{aligned}
\end{equation}
where $\mat{A}$, $\mat{B}$ and $\mat{C}$ are lazy matrices
and $\vec{x}$ is a vector stored in memory.
\begin{figure}
	\centering
	\includegraphics{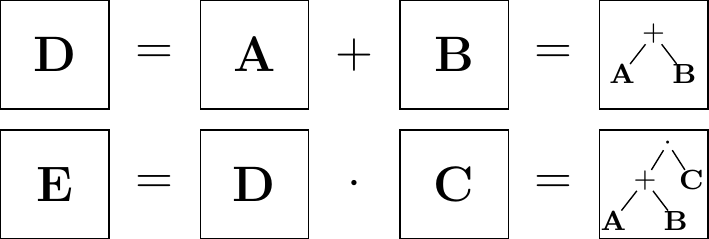}
	\caption[Examples for lazy matrix expression trees]
	{
		Examples for lazy matrix expression trees.
		The upper represents the instruction
		$\mat{D} = \mat{A} + \mat{B}$
		and the lower the multiplication of the result $\mat{D}$
		with $\mat{C}$.
	}
	\label{fig:LazyMatrixExpressionTree}
\end{figure}
The first two lines do not give rise to any computation.
They only build an expression tree in the returned
lazy matrix $\mat{E}$, as illustrated in figure \ref{fig:LazyMatrixExpressionTree}.
The final instruction is a matrix-vector product with the stored vector $\vec{x}$,
and the actual result should be returned in the vector $\vec{y}$.
This triggers the complete expression tree to be worked upon in appropriate order,
such that 
the expression
\begin{equation}
	\vec{y} = \left( \mat{A} + \mat{B} \right) \mat{C} \vec{x}
	\label{eqn:LazyMa}
\end{equation}
is evaluated at once at this very instance.
At this point, the full
expression could be simplified, and the most efficient evaluation order
could be chosen. This is, however, not implemented yet.

Due to lazy evaluation, we can thus build complicated expressions
from familiar matrix operations.
This allows to view the lazy matrix framework as a domain-specific language
for \contraction-based algorithms,
which makes working with \contraction expressions feel like working
with actual matrices.
Note, however, that the lazy matrices generalise real matrix
as they also allow non-linear transformations of
a vector to be represented.

Inside \lazyten, various kinds of lazy matrices,
as well as lazy and stored objects, can be combined transparently~\cite{Herbst2018Phd}.
Similarly, \lazyten provides
high-level interfaces for solving linear or eigenproblems~\cite{Herbst2018Phd}
where the involved matrices may be stored or lazy.
The call passes through a branching layer,
which inspects the matrix structure
and accordingly selects one of the available
third-party linear algebra back ends for solving the problem.
As a result, algorithms programmed using \lazyten
may be called with both lazy and stored matrices,
and the solvers will be automatically chosen to match
the change in matrix structure.
The user, however, remains in full control:
By providing appropriate parameters all choices
made automatically by \lazyten can be overwritten, as can
the parameters passed to the underlying solvers.

Overall, \lazyten provides intuitive, high-level syntax
for \contraction-based methods in the form of lazy matrices.
The library allows algorithm code to be written only once,
but to stay flexible.
For example,
one may
adapt to modern hardware trends or
to the deviating numerical requirements
imposed by a different basis function type
simply
by changing the implementation of the lazy matrices
passed to the algorithm code.
A more in-depth discussion of \lazyten can be found in
reference \onlinecite{Herbst2018Phd}.

%
%

\subsection{Self-consistent field methods and integral interface}
\label{sec:GscfGint}

The lazy matrices of \lazyten are constructed to
be used as a high-level language
for implementing basis-type-independent, contraction-based
SCF algorithms.\cite{Herbst2018Phd}
For example, the linear algebra interfaces
of \lazyten can be employed in the coefficient-update step
making implicit use of the automatic switching
between dense and iterative diagonalisation methods.
The Fock update may be implemented building on top of
a similar function from \lazyten, namely by altering
the coefficient matrix the Fock expression currently refers to.

All the \SCF algorithms in \gscf are implemented in this way:
as solvers for a non-linear eigenproblem
defined by an input lazy matrix,
which represents the SCF problem under study.
Since the algorithms only see the final \contraction expression
and its update function,
\gscf is self-contained and
may be applied to \emph{any} non-linear eigenproblem
with a structure similar to the \HF minimisation problem.
This is desirable,
because quite a few electronic structure theory methods
can be thought of as modifications of the \HF problem.
Examples include the Kohn-Sham matrix
arising in the usual density-functional theory~(\DFT) treatments
or additional terms in the problem matrix,
arising from modelling an external field,
or correction terms due to embedding.

The lazy Fock matrix object describing the problem to be solved
is prepared by the upper \molsturm layer
based on the electronic structure method chosen by the user.
For example, \HF would be the sum of four lazy matrices,
which represent the
kinetic, nuclear attraction, Coulomb, and exchange matrix~\cite{Helgaker2013}.
Similarly one would add an exchange-correlation term for \DFT,
or other terms such as an embedding operator.
The latter methods are not yet available in \molsturm, however.

The individual terms of the Fock matrix are obtained from \gint,
which acts as broker, presenting a common interface for all
basis function types and third-party integral back end libraries
to the rest of the \molsturm ecosystem.
On calculation start, \molsturm will take the discretisation parameters
supplied by the user and hand them over to \gint,
which --- based on these parameters ---
sets up the selected integral back end library
and returns a collection of lazy matrix integral objects.
For each basis type and back end, the interface of the returned objects
will thus look alike, since they are all lazy matrices.
On call to their respective \contraction expressions, however,
the required computations will be invoked in the previously selected
integral back end.
\gint itself does not implement any routine
for computing integral values at all,
it just transparently redirects the requests.
Notice, that the precise kind of parameters needed by \gint
to setup the back end may well vary from discretisation to discretisation.
For example, a Coulomb Sturmian basis requires the Sturmian exponent $k$
and the selection of $(n, l, m)$-triples of the basis functions~\cite{Herbst2018Phd}
whereas a contracted Gaussian basis requires the list of angular momentum,
exponents and contraction coefficients~\cite{Hehre1969,Jensen2013,Hill2013}.

At the moment cGTO and CS integrals
are 
the only ones supported in \gint.
For each of these, at least two different implementations are available, however.
Adding more back end libraries or basis function types is rather easy,
since one only needs to implement a collection of lazy matrices,
where the \contraction expressions initiates the appropriate
integral computations in the back end.
This collection then needs to be registered as a valid basis type
to \gint to make it available to the upper layers.
For example, 
preliminary support for the contracted Gaussian library
\libcint~\cite{Sun2015} was added with just two days of work.
Notice, that \gint is designed to allow all of this to be achieved
without changing a single line of code inside \gint itself,
since the call to the registration function can happen dynamically at runtime.
So one can implement a new integral back end in a separate shared library
and add it in a plug-in fashion without recompiling \molsturm.

To summarise,
by means of the lazy matrices of \lazyten,
the responsibility for the \HF problem
has been split between three different,
well-abstracted modules:
\gint, which provides the interface to the integrals and selects the discretisation,
\molsturm, which builds the lazy matrix expression of the problem to be solved
and \gscf which uses this expression to solve the \SCF problem
in a basis-type-independent manner.

%
%

\subsection{\python interface module}
\label{sec:python}
The topmost layer of \molsturm
is the ``\molsturm'' \python module,
providing the user interface of the package.
This layer assists with setting up a calculation,
drives the \SCF procedure in \gscf
and returns the converged results to the user.
We chose the scripting language \python to implement the majority
of this interface layer and especially the interface itself.

Our reasoning is related to the arguments of the \pyscf authors~\cite{Sun2017}
discussed in section \ref{sec:related},
namely we wanted to
avoid inventing yet another ``input format'' and ``output format''
for quantum-chemical calculations.
Instead, calculations in \molsturm can be initiated cleanly and flexibly
directly from a host \python script,
which can additionally be used for subsequent analysis.
This not only implies that all of \python and its libraries
are available for the calculation setup and analysis,
but also
that no explicit parsing of program output is required
for analysing the results.
This lowers the barriers for people
who are new to the field,
since they do not need
to learn both how to write input files that define calculations,
as well as the syntax of a scripting language for parsing results.
More subtly, the output formats
of quantum-chemistry programs change from time to time,
breaking parser scripts or --- even worse ---
producing wrong results without any notice.
This is a common problem in the current practice of computational chemistry.

In contrast to this, the \SCF results in \molsturm are returned to the host \python
environment through an interface
built
on \numpy arrays.
These have become the \textit{de facto} standard
for storing and manipulating matrices or tensors in \python.
All \python packages
that are commonly used for plotting or data analysis,
such as matplotlib~\cite{Matplotlib}, scipy~\cite{Walt2011,scipyWeb},
or pandas~\cite{pandasWeb},
use \numpy arrays in their interfaces.
Consequently, a complete computational procedure
may be orchestrated from a single \python script, which contains
all parameters influencing setup, calculation, analysis
and all decisions taken for presenting the data in plots or tables.
Such a script serves as automatic documentation for the full procedure
and allows others to reproduce the presented plots or tables without effort:
All it takes is to re-run the script.

\emph{All} parameters for \gint, the \SCF algorithms of \gscf,
as well as the linear solvers from \lazyten are made available
through the \python interface.
By changing these, the user may directly influence, e.g., the algorithms
chosen by \lazyten for diagonalisation,
or how \gscf switches between 
\SCF solvers.
This is particularly handy during method development,
where one may run \molsturm from an interactive IPython~\cite{IPython} shell
and use these parameters to
control 
the progress of a calculation.
In that way one can
check assertions about intermediate results
or visualise such graphically with matplotlib~\cite{Matplotlib}.
This greatly reduces the feedback loop for small calculations,
\eg during debugging.

Interactive analysis of larger calculations is facilitated
by archiving functionality in \molsturm.
The \SCF results
may be stored either in
YAML~\cite{YAML} or HDF5~\cite{HDF5Manual} format.
In this way large calculations an be performed in advance
over night or on an high-performance computing~(HPC) system,
then archived and transferred to the workstation.
Here, the archive may be loaded in an interactive shell,
restoring the full state of the calculation
as if it had been performed locally.
Next to the \SCF results,
\molsturm's archived state contains
the precise set
of input parameters which were used to obtain the stored results.
These are \emph{not} the parameters provided by the user to start the calculation,
but the post-processed parameters which were actually used
by the lower layers, including e.g.~default values.
This helps make the archive self-documenting,
and simplifies
setting up a refined calculation building on top of the already obtained results.

Our \numpy-based interface has already proven to be helpful 
for linking \molsturm to other third-party quantum chemistry codes:
It allowed us to link \molsturm to
the \python interfaces of
\pyscf~\cite{Sun2017}, as well as \adcman~\cite{Wormit2014}
in only a few days. 
As a result, FCI as well as 
calculations
for computing excited states by the algebraic diagrammatic construction~(ADC) scheme%
~\cite{Schirmer1982,Trofimov1999}
may be started on top of \molsturm's \SCF
using the respective aforementioned packages.
By way of interface generators like SWIG~\cite{swigWeb},
\numpy arrays can be automatically converted to plain
\ccc arrays, such that third-party packages
consisting only of low-level \cpp, \ccc or \fortran code
can be linked to \molsturm in the future.

We see that the
interface of molsturm
not only facilitates rapid development of new algorithms,
but also relieves one from the need to re-invent the wheel,
\ie to implement standard quantum-chemical methods over
and over for each new basis function type.
Instead, existing functionality in external packages can be quickly
leveraged for one's own purposes.
The aspects described in this section will be demonstrated
with practical examples in the next section.


\section{Examples}
\label{sec:examples}

In this section we present two examples
that demonstrate how the \python interface of \molsturm
can be combined with existing features of \python
in order to analyse results or to extend the capabilities of \molsturm.
We concentrate our discussion on molecular computations
with contracted Gaussian basis sets.
It should be stressed again, however,
that due to the basis-function independent nature of \molsturm,
the procedures outlined in the scripts could be easily performed
with other types of basis functions as well.

In fact, the design of \molsturm
assures that the discretisation details can be selected
at a high level,
without affecting the code
that performs the actual computation and analysis.
Section \ref{sec:ex:geo} gives an example
for which the choice of the basis type
is made in the \texttt{main} function of the script.
This ensures that a script performing a particular modelling task
can be easily used as a template for
a systematic study of the effect 
of changing basis function type or integral implementation:
All it takes is to
iterate over the appropriate list of discretisation
parameters and call the calculation for each instance.
This greatly simplifies testing a novel basis function type,
which has just become available in \gint,
as well as comparing it to existing methods
subject to the test case provided by a script.

\subsection{Coupled-cluster doubles~(CCD)}
\label{sec:ex:ccd}

\begin{figure*}
	\includegraphics[scale=0.9]{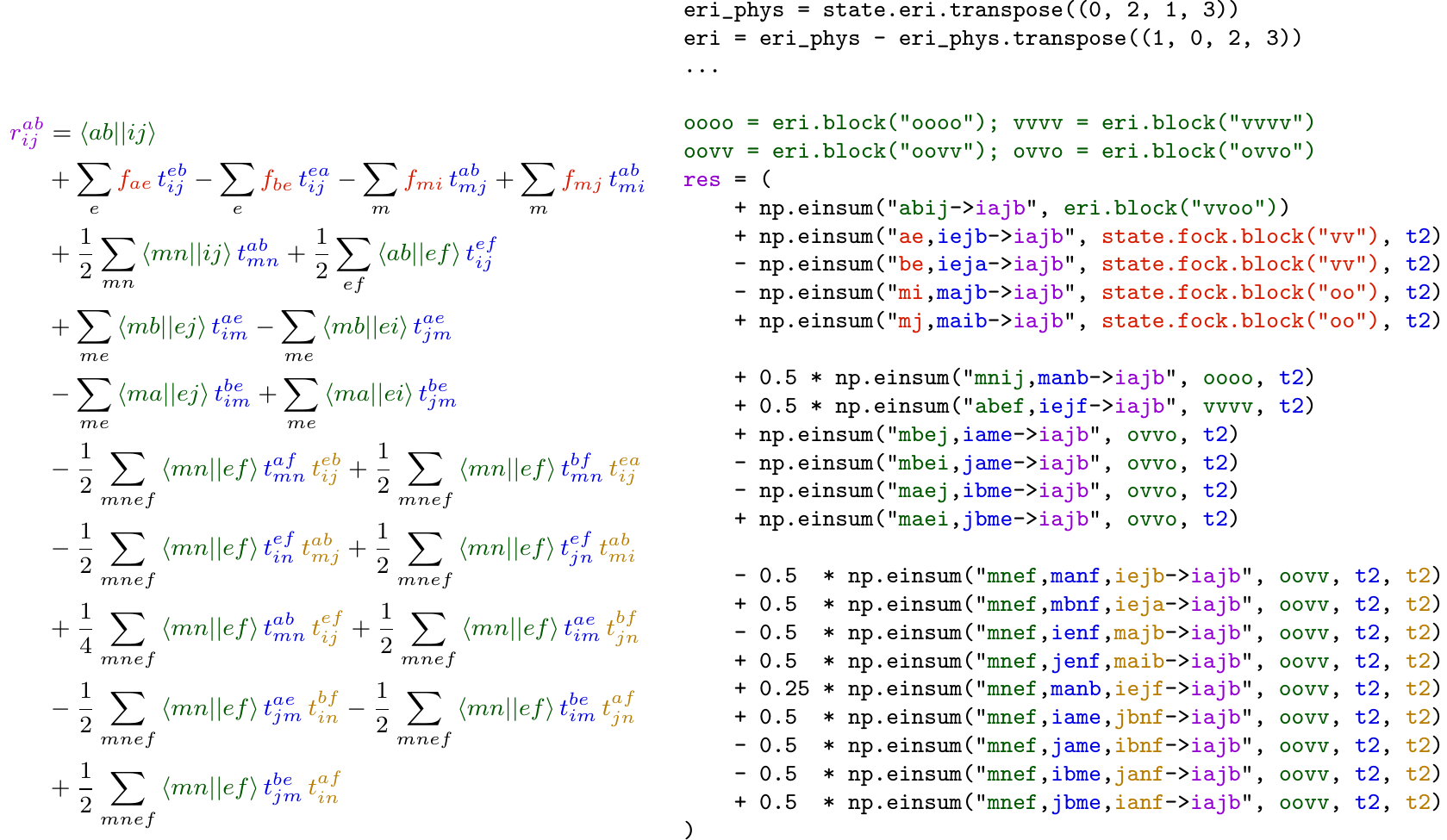}
	\caption{Equation for the coupled-cluster doubles~(CCD)
		residual~\cite{Bartlett1978}
		on the left and excerpt of a CCD implementation
		using \molsturm and \numpy on the right.
		Equivalent quantities are highlighted in the same colour.
		The first two lines of code show the computation of the
		antisymmetrised electron repulsion integrals
		from the \texttt{state.eri} object obtainable from \molsturm,
		which is carried out once at the beginning of the algorithm.
		The remaining lines compute the residual for
		a particular $T_2$ amplitude stored in the tensor object
		\texttt{t2}.
	}
	\label{fig:codeCCD}
\end{figure*}

This example shows how one can extend \molsturm with novel methods
using its high-level \python interface together with standard functionality
from \python/\numpy~\cite{Walt2011,scipyWeb}.

Even though \molsturm right now neither offers
any coupled-cluster method nor
an interface to any third-party coupled-cluster code,
we managed to implement a simple, working
coupled-cluster doubles~(CCD)~\cite{Hurley1976,Bartlett1978} algorithm
in only about 100 lines of code and about two days of work,
including the time needed for research about the method
and the computational procedures.
The most relevant part of the implementation,
namely computing the CCD residual for the current guess
of the $T_2$ amplitudes $t_{ij}^{ab}$,
is shown towards the right of figure \ref{fig:codeCCD},
side-by-side with the expression of the CCD residual~\cite{Bartlett1978}.
The full CCD code is available as an example
in the file
\url{examples/state_interface/coupled_cluster_doubles.py}
of the \molsturm repository~\cite{molsturmWeb}.
We follow the standard procedure
of employing a quasi-Newton minimisation
of the CCD residual with respect to the $T_2$ amplitudes
using the orbital energy differences as an approximate Jacobian%
~\cite{Bartlett1978,Helgaker2013}.
The guess for the $T_2$ amplitudes is taken
from second order Møller-Plesset perturbation theory.

The expression of the CCD residual $r_{ij}^{ab}$
requires the evaluation of a sequence of tensor contractions
involving the Fock matrix in the molecular orbital basis, $f_{pq}$,
the antisymmetrised electron-repulsion integrals, $\langle pq||rs \rangle$,
as well as the current guess for the $T_2$ amplitudes, $t_{ij}^{ab}$.
As usual, we employ the convention that indices $i,j,k,l,\ldots$ refer
to occupied orbitals, indices $a,b,c,d,\ldots$
to virtual (\ie unoccupied) orbitals, 
and indices $p,q,r,s,\ldots$ to either kind of orbitals.

The \python implementation (right-hand side of figure \ref{fig:codeCCD})
computes those contractions.
For this it employs the data structures \molsturm provides
in the \texttt{state} object, which is returned by the SCF procedure.
Our code uses chemists' indexing convention
in the electron-repulsion integrals object \texttt{state.eri}.
The CCD equations, however, are written using
the antisymmetrised electron-repulsion integrals $\langle pq||rs \rangle$.
Therefore the first two lines of the code of figure \ref{fig:codeCCD}
are executed once to perform the antisymmetrisation.
The subsequent lines are executed once per CCD iteration and compute the residual tensor \texttt{res}
by contracting the relevant blocks of the Fock matrix \texttt{state.fock},
the \texttt{eri} object and the $T_2$ amplitudes contained in \texttt{t2}.
This is implemented using the \texttt{einsum} method from \numpy,
which performs tensor contractions
expressed in Einstein summation convention.
Note how the interplay of \numpy
with the data structures \molsturm results in a
strikingly close resemblance of implementation and actual equation.

The \texttt{state} object
provides access to more quantities from the SCF procedure
than just the Fock matrix and the repulsion integrals.
Individual terms of the Fock matrix or
quantities like the overlap matrix in terms of the underlying
discretisation basis functions may be obtained as well.
We provide this data as \numpy arrays
extended with extra functionality
to simplify implementation of Post-HF quantum-chemical methods,
such that the user can employ the SCF results freely and flexibly
within the \python ecosystem.
Coupled with the basis-function independence of \molsturm's
SCF this allows for rapid development and systematic investigation
of Post-HF methods based on arbitrary basis functions.

At the moment we make no efforts to exploit symmetry or parallelise
the computation of the tensor contractions
shown in the script of figure \ref{fig:codeCCD}.
For this reason, such implementations are not suitable for
real-world applications.
Nevertheless, the script presented in figure \ref{fig:codeCCD}
may be used for CCD calculations of small molecules with small basis sets.
For example, an \ce{O2} 6-31G~\cite{Hehre1972} calculation on a recent laptop took
about an hour to converge up to a residual $l_\infty$-norm of $10^{-4}$.
For investigating new methods on top of the \molsturm framework,
or to provide a flexible playground for teaching Post-HF methods to students,
such scripts are therefore still well-suited.

\subsection{Gradient-free geometry optimisation}
\label{sec:ex:geo}

\begin{figure}
	\centering
	\includegraphics[width=0.48\textwidth]{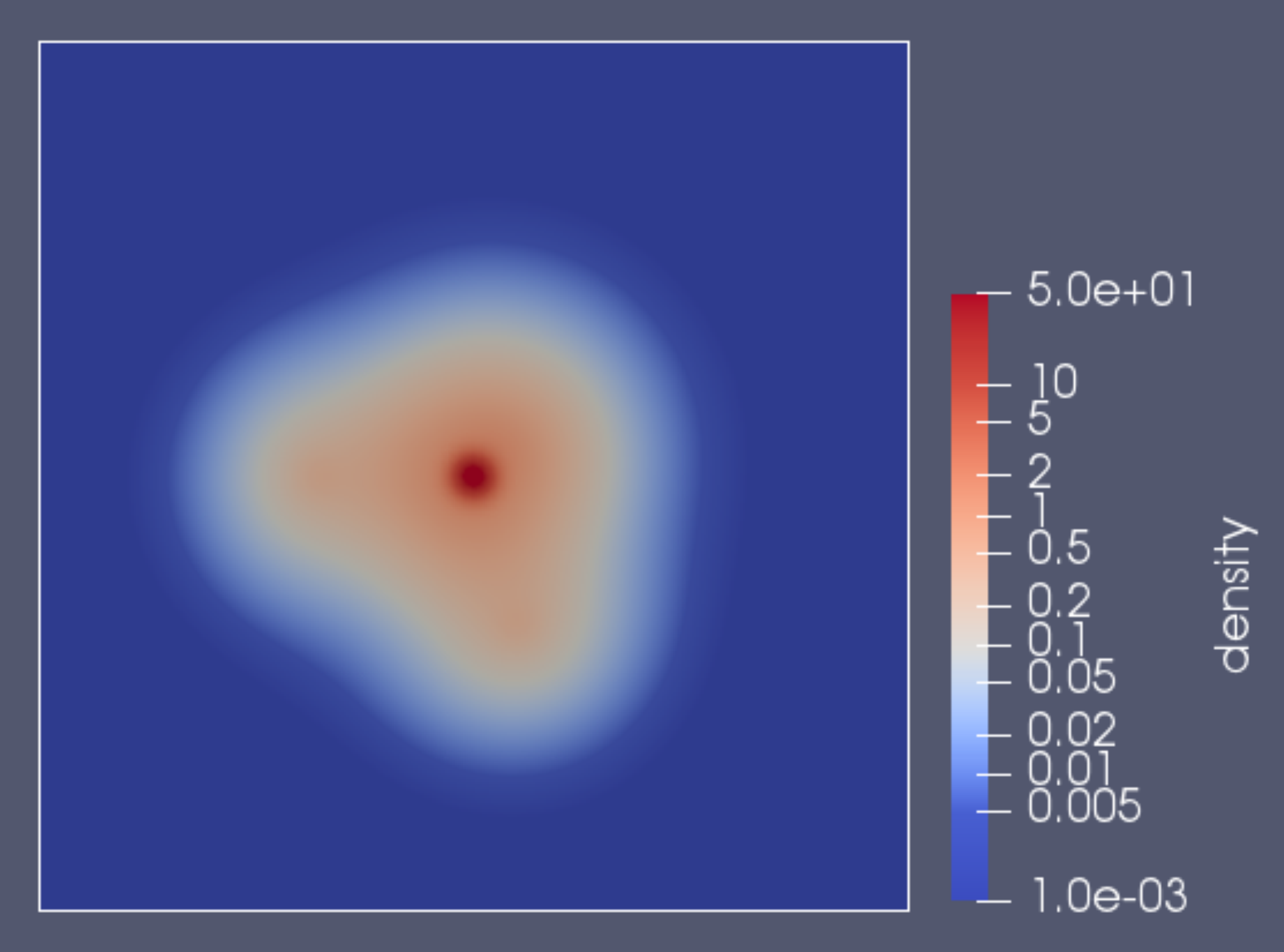}
	\caption{Density plot of the final optimised \ce{H2O}
	Hartree-Fock geometry with a
	\ce{O-H} bond length of \unit[0.95046]{\AA} and
	a \ce{H-O-H} bond angle of $106.35^\circ$.
	A geometry optimisation in ORCA~\cite{ORCA}
	employing the same basis set
	agrees with this result within the convergence tolerance of $10^{-5}$.
	}
	\label{fig:OptimalGeometryWater}
\end{figure}

\begin{figure}
	\includegraphics[scale=0.95]{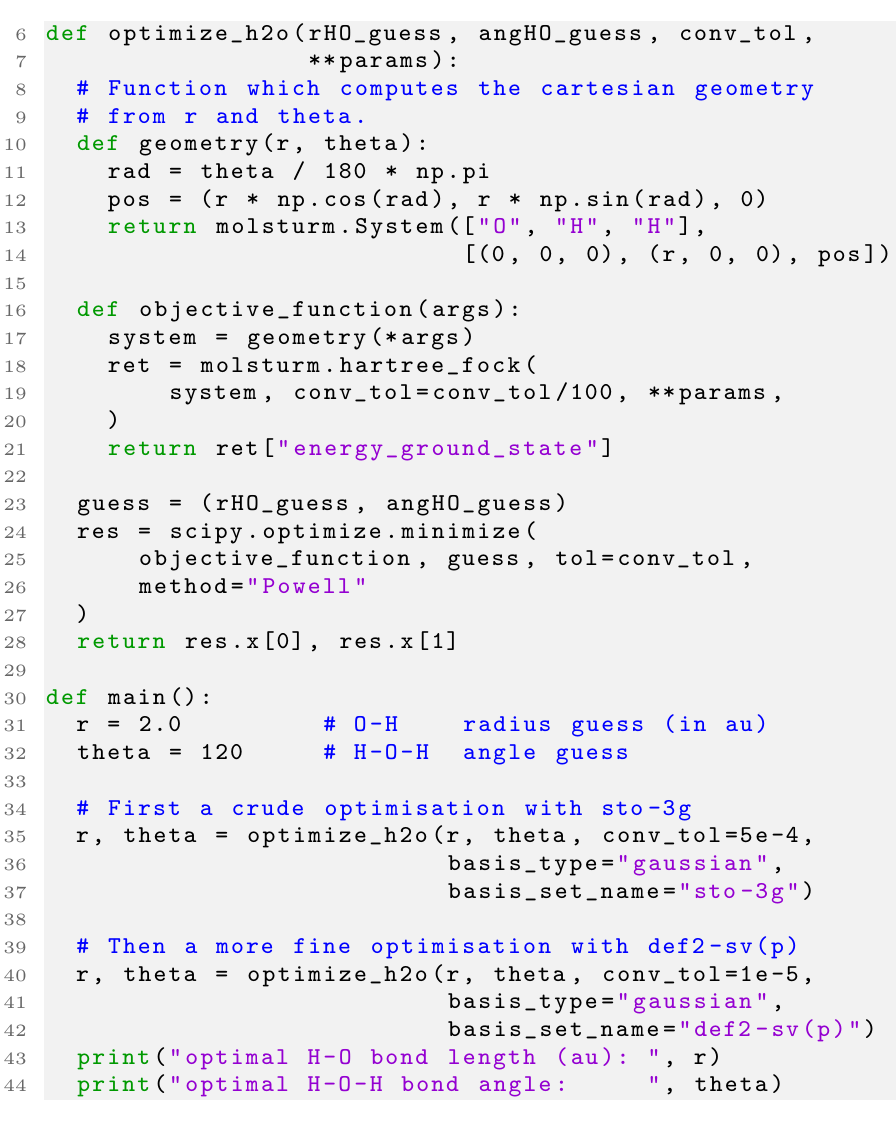}
	\caption{
		Example for performing a gradient-free optimisation
		using Powell's method~\cite{Powell1964,Press1992} and \molsturm.
		\python \texttt{import} statements at the top of the script
		and the explicit call to \texttt{main} are skipped.
	}
	\label{fig:codeGeoOpt}
\end{figure}
\newcommand{\lgbasone}{36\xspace}
\newcommand{\lgbastwo}{41\xspace}

In order to make a novel basis function type properly accessible
to the full range of quantum-chemical methods,
a daunting amount of integral routines and computational procedures
need to be implemented.
For assessing the usefulness of a new discretisation method it is, however,
important to be able to quickly investigate its performance with respect to
as many problems as possible.
Undoubtedly, a very important application of computational chemistry
is structure prediction, \ie geometry optimisation.
Performing such calculations requires
the appropriate integral derivatives for the chosen basis function type.
Since implementing these in the integral library can be as difficult as implementing the integrals
required for the SCF scheme itself,
one would much rather skip this step at first, and concentrate
only on what is required for the SCF.

This example extends \molsturm with a gradient-free geometry-optimisation procedure,
implemented with building blocks readily available from \python.
This sidesteps the need for nuclear derivatives on the side of the integral
library and facilitates simple structure optimisations, even without nuclear gradients
--- neither analytical nor numerical.

Figure \ref{fig:codeGeoOpt} shows the script,
which performs a geometry optimisation of water
based on Powell's gradient-free optimisation algorithm~\cite{Powell1964,Press1992}
as implemented in the \texttt{scipy} library%
~\cite{Walt2011,scipyWeb}.
The optimal structure is found in a two-step procedure.
First, a cheap \mbox{STO-3G}~\cite{Hehre1969} basis set is used to obtain
a reasonable guess.
Then, the final geometry is found
by minimising to a lower convergence threshold in the more costly
\mbox{def2-SV(P)}~\cite{Weigend2005} basis.

The time required to code the script was only about 30 minutes,
showing the great power of a flexible design.
Nevertheless, convergence to the equilibrium geometry shown
in figure \ref{fig:OptimalGeometryWater}
was achieved in a couple of minutes.
In line with what was discussed above,
a novel basis function type,
for which one just implemented the \SCF integrals in \gint,
can be directly used for geometry optimisations.
Only the discretisation parameters need to be changed,
in lines \lgbasone and \lgbastwo of the outermost \texttt{main} function.


\section{Current state and future of \molsturm}
\label{sec:state}

After about two years of development, \molsturm
allows to solve the Hartree-Fock~(HF) equations
basis-function independently,
following a  \contraction-based, self-consistent field~(SCF) ansatz.
All aspects of the calculation, including the diagonalisation algorithm
and the basis function type of the discretisation,
may be fully controlled via a \python module.
This module integrates well into the existing \python ecosystem,
simplifying repetitive calculations as well as
analysis of obtained results.

At present, \molsturm's integral library \gint
supports
calculations employing either Coulomb Sturmians or contracted Gaussian
basis functions, both in multiple implementations.
For contracted Gaussians, the third-party
\libint~\cite{Libint2,Libint2_231} or \libcint~\cite{Sun2015} libraries can be used,
and Coulomb Sturmians are available via
our own \sturmint~\cite{sturmintWeb} library.
In the future we plan to add support for further basis function types
in \gint and \molsturm, in particular molecular and generalised Sturmians%
~\cite{Avery2012,Avery2013,Avery2015,Avery2017}.

Via \gscf, multiple SCF schemes are available,
namely Roothaan's repeated diagonalisation~\cite{Roothaan1951},
Pulay's commutator direct inversion of the iterative subspace~(DIIS)~\cite{Pulay1982}
as well as the truncated optimal damping algorithm~(tODA)~\cite{Herbst2018Phd}, 
an approximation of the optimal damping algorithm~\cite{Cances2000a},
which is more suitable for the contraction-based interface of \gscf.
During the SCF procedure, \molsturm automatically switches between
the available schemes,
trying to balance convergence rate and expense of the
individual algorithms.

Once an SCF computation has finished, the results can be archived
in either in YAML~\cite{YAML} plain text or in HDF5~\cite{HDF5Manual} binary files.
Such an archive not only contains the full final state of the calculation
but also the precise parameters which were used in the SCF procedure,
making the archive file self-documenting.

For treating electron correlation,
\molsturm only implements
second order Møller-Plesset perturbation theory~(MP2).
Further methods, however, can be easily called
via interfaces to third-party libraries.
Full configuration interaction~(FCI) is available via \pyscf,
and a range of excited states-methods
based on the algebraic diagrammatic construction~(ADC) scheme
via \adcman~\cite{Wormit2014}, namely
ADC(1), ADC(2), ADC(2)-x~\cite{Schirmer1982}
and ADC(3)~\cite{Trofimov1999}.

The extension of \molsturm to other methods or packages
is easily accomplished by \molsturm's
\python interface, see section \ref{sec:examples}.
Along these lines, closer integration with projects
such as \pyscf~\cite{Sun2017}, \psifour~\cite{Parrish2017}
or \psifnp~\cite{Smith2018}
could be promising,
since these already provide high-level \python interfaces
to many state-of-the-art Post-HF methods.
In this way,
configuration-interaction, coupled-cluster,
multi-configurational self-consistent field
or density matrix renormalisation group approaches
could be used directly from \molsturm's \SCF.
With manageable development time,
these methods would thus become available
for all basis function types implemented in \gint.
Similarly,
the extension of \molsturm's \SCF towards Kohn-Sham density-functional theory
is possible employing third-party libraries
such as
\texttt{libxc}~\cite{Lehtola2018} or \texttt{xcfun}~\cite{xcfun}
for computing the required exchange-correlation integrals.

As discussed in section \ref{sec:Lazyten},
a contraction expression inside \molsturm's \SCF is evaluated
whenever the Fock matrix is applied to a trial vector.
This proceeds by working on the expression tree,
which represents the Fock matrix.
In \lazyten, the corresponding computations
are right now neither parallelised,
nor are symmetries or repetitive terms in the expression tree
exploited.
This currently limits the applicability of \molsturm's \contraction-based \SCF
to small basis set sizes.
Both automatic parallelisation of linear algebra expressions
and finding optimal evaluation schemes for expression trees,
is ongoing research%
~\cite{Baumgartner2005,Solomonik2014,%
Peise2015,Calvin2015,Xerus,Kristensen2016array,%
Kristensen2016streaming,Libtensor}, however.
By integrating such efforts into \lazyten
a direct improvement of \molsturm's \SCF could be achieved
without changing any other code.


\section{Discussion and conclusion}
\label{sec:conclusion}

Implementation of quantum-chemical methods
using novel types of basis functions often requires
unusual numerical techniques as well.
Implementing these into existing quantum chemistry packages
can be a large task,
as these are highly optimised for the methods they already accomodate,
and are typically not flexible enough to meet other requirements.

The \molsturm framework presented here
tries to fill this gap
by providing a light-weight package designed with a range of different
basis functions in mind.
The key ingredient to reach the necessary flexibility
is a \contraction-based self-consistent field~(SCF) scheme,
which we employ for solving the Hartree-Fock problem.
In a \contraction-based ansatz,
the numerical algorithms are formulated without requiring any
explicit reference to the Fock matrix memory.
Instead, the SCF iterations are driven
by contractions of the Fock matrix with other vectors.
The details how this matrix-vector product is computed
can be varied flexibly, matching the numerical properties
of the discretisation at hand.
In this way, we have reached a design where
the code for
the SCF algorithm is separated from the code performing the linear algebra
computation.
Thus, changes to the SCF scheme can be made without
affecting other parts of \molsturm
and the SCF code itself becomes basis-function independent.

On top of that, the interfaces
of our SCF are easy-to-use and readily extensible.
This allows quick incorporation of functionality
of third-party packages and extensions of \molsturm in ways we as the authors
would have never thought of.
Right now, \molsturm may be used to perform
calculations based on contracted Gaussians~\cite{Hehre1969} --- using
the integral libraries \libint~\cite{Libint2_231,Libint2}
or \libcint~\cite{Sun2015} --- and based on
Coulomb Sturmians~\cite{Shull1959,Rotenberg1970} ---
using \sturmint~\cite{sturmintWeb}.
Selected Post-HF methods from \pyscf~\cite{Sun2017} as well as
excited states methods from \adcman~\cite{Wormit2014} are available on top.
Extending the set of basis function types available
inside \molsturm can be achieved in a plug-and-play fashion,
namely by implementing a single, well-defined interface class in
our integral library \gint.
Thereafter such basis functions are available for the full \molsturm ecosystem
including the Post-HF methods
provided by the third-party libraries mentioned above.

The abilities of \molsturm have been demonstrated
by two practical examples with particular emphasis
on the way our \python interface integrates with existing \python packages.
We showed how to aid repetitive calculations,
implement novel quantum-chemical methods
or rapidly amend functionality in a preliminary way,
where a proper implementation would be much more involved.
We hinted how
systematic comparisons with established basis functions
as well as subsequent graphical analysis
is convenient to perform by the means of our
readily scriptable interface.
We hope that in this manner, \molsturm
will be a useful package to rapidly try novel basis function types
and get a feeling for their range of applicability.


\section{Acknowledgments}
The authors express their thanks to \ademp and \maxsch
for many fruitful discussions during the preparation of the work.
Michael F. Herbst gratefully acknowledges funding by the
Heidelberg Graduate School of Mathematical and Computational
Methods for the Sciences (GSC220).
Last but not least the authors wish to
commemorate their former collaborator,
supervisor and friend Dr. Michael Wormit,
whose ideas have survived in this project as well as in countless others.


\bibliography{literature}
\end{document}